# Time-resolved high-harmonic spectroscopy of ultrafast photo-isomerization dynamics

Keisuke Kaneshima,* Yuki Ninota, and Taro Sekikawa

*Division of Applied Physics, Hokkaido University, Sapporo, Hokkaido 060-8628, Japan*

*kaneshima@eng.hokudai.ac.jp

**Abstract**

We report the first study of time-resolved high-harmonic spectroscopy (TR-HHS) of a bond-making chemical reaction. We investigate the transient change of the high harmonic signal from 1,3-cyclohexadiene (CHD), which undergoes ring-opening and isomerizes to 1,3,5-hexatriene (HT) upon photoexcitation. By associating the change of the harmonic yield with the changes of the ionization energy and vibrational frequency of the molecule due to the isomerization, we find that the electronic excited state of CHD created *via* two-photon absorption of 3.1 eV photons relaxes almost completely within 80 fs to the electronic ground state of CHD with vibrational excitation. Subsequently, the molecule isomerizes abruptly to HT, i.e., ring-opening occurs, around 400 fs after the excitation. The present results demonstrate that TR-HHS, which can track both the electronic and the nuclear dynamics, is a powerful tool for unveiling ultrafast photo-chemical reactions.

**Body**

Measuring and understanding ultrafast dynamics in matter has always been a prime research goal. Recent developments of ultrashort-pulse light sources have allowed us to study electronic dynamics at the attosecond and angstrom scales, and progress has been made via high-harmonic generation (HHG), which provides attosecond pulses in the extreme ultraviolet (EUV) and enables the development of time-resolved techniques in the attosecond regime[1–4].

Not only as an ultrashort EUV light source, HHG process itself can also be used to investigate the state of the generating medium. HHG can be understood in the following sequence[5,6]: (i) A strong laser field tears a valence electron away from an atom or molecule through tunnel ionization. (ii) The free electron is accelerated by the laser electric field and then is pulled back to its parent ion when the sign of the electric field reverses. (iii) The laser-driven electron recollides and recombines with its parent ion. As a result of recombination, the kinetic energy of the recolliding electron plus the binding energy are released through coherent radiation, i.e., HHG, which contains information about the electron-ion interaction. Hence, one can retrieve the electronic state of the generating medium from the amplitudes and phases of the high-harmonic spectra. This is known as high-harmonic spectroscopy (HHS)[7,8]. On account of the strong nonlinearity of the ionization process, HHS of molecules can sensitively and selectively probe the highest occupied molecular orbitals (HOMOs), which are of prime importance for understanding chemical reactions.

HHS has been used to study the ground-state electronic wavefunctions of gas-phase molecules[9–15]. In these studies, the molecular-alignment technique *via* rotational excitation[16] has been exploited to characterize the HHG from molecules as a function of their spatial orientations. Ultrafast molecular dynamics triggered by vibrational or electronic excitation have also been investigated via time-resolved HHS (TR-HHS), where the high-harmonic signals are monitored as a function of the pump-probe delay. TR-HHS has been used to study the vibrational dynamics of $SF_6$[17] and $N_2O_4$[18] and photo-dissociation dynamics of $Br_2$[19], $NO_2$[20], $CH_3I$, and $CF_3I$[21]. These experiments have shown that the HHG process is sensitive to both the valence electronic and the vibrational states of the molecules. Since chemical reactions result from the coupled dynamics of valence electrons and nuclei, the ability to simultaneously monitor the electronic and nuclear dynamics of molecules makes TR-HHS ideal for probing ultrafast chemical reactions.

However, applications of TR-HHS have previously been limited to photo-dissociation reactions of small molecules[19–21]. The TR-HHS of more complicated reactions like the concerted breakage and formation of bonds initiated by photoexcitation remains a challenge. This limited application mainly stems from the difficulty of observing high-harmonic signals from photoreactive organic molecules. Since these molecules are in a condensed phase at room temperature, due to their polarity and large molecular weights, their vapor pressure is not high enough to observe high-harmonic signals[8].

Here, we overcome this difficulty and demonstrate TR-HHS of ultrafast photo-isomerization dynamics of 1,3-cyclohexadiene (CHD), $C_6H_8$. This is a significant advancement, which makes TR-HHS a more versatile and practical tool for studying ultrafast chemical reactions.

CHD comprises a hydrocarbon ring that exists in the liquid phase at room temperature. It undergoes ring-opening and isomerizes to 1,3,5-hexatriene (HT) upon photoexcitation[22,23] (Fig. 1). As a prototypical example of an electrocyclic reaction, this reaction plays an important role in the understanding of a large number of organic reactions, including the photo-induced formation of provitamin D[24,25]. It is also used as a touchstone for novel ultrafast spectroscopic techniques, including x-ray fragmentation[26] and x-ray scattering[27] with femtosecond x-ray pulses from a free-electron laser, photoelectron spectroscopy with single-order harmonic pulses[28,29], and near edge x-ray absorption fine structure at the carbon K-absorption edge with soft-x-ray high harmonics[30].

The ring-opening dynamics of CHD after excitation *via* one-photon absorption (1PA) with a photon energy larger than 4.5 eV is one of the most actively studied dynamics in time-resolved spectroscopy[26,27,29–40]. Most studies have found that the ring-opening step in the isomerization process is completed in less than 200 fs after photoexcitation. In contrast, we have investigated the ring-opening of CHD *via* two-photon absorption (2PA) using 3.1 eV photons for time-resolved photoelectron spectroscopy (TR-PES). We used the isolated single-order harmonic at 29.5 eV[28] selected by a time-delay-compensated monochromator[41–44]. In that study, we reported that the ring opens at around 500 fs, which is much slower than in the 1PA cases. Accordingly, we have been motivated to obtain deeper insight into the ring-opening dynamics caused by 2PA, using TR-HHS. We expect TR-PES and TR-HHS will provide complementary information because the former is sensitive to deep molecular orbitals and the latter is sensitive to the shallowest orbital.
In addition, although CHD can adopt three distinct conformations after ring-opening, i.e., cZc-, tZt-, and cZt-HT (Fig. 1), the structural isomerization dynamics of the HTs after ring-opening have rarely been discussed to date. In this study, we also endeavor to clarify the structural-isomerization dynamics of the HTs using TR-HHS.

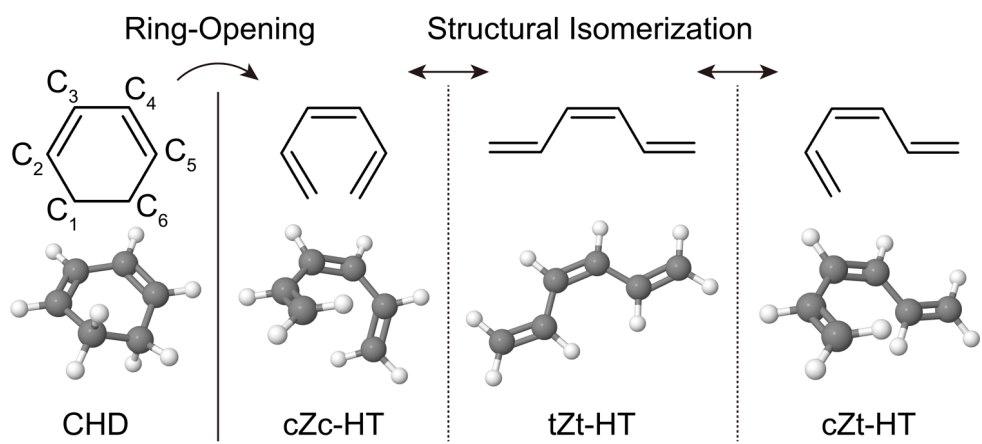

**Fig. 1** Structural formulas and corresponding three-dimensional representations of CHD and its isomers.

# Results

In Fig. 2a, we show the experimentally observed 19th harmonic yield as a function of pump-probe delay (see Methods section). The dip in the harmonic yield at zero delay is a coherent artifact due to the overlap of the pump and probe. A polarization component orthogonal to the probe polarization is introduced by the pump, and it reduces the recombination probability in the HHG process. The full-width at half-minimum of the dip was ~80 fs, providing a high-order cross-correlation time between the pump and probe pulses. After excitation, we observed steep decreases in the harmonic yield around 400 and 1100 fs with a gradual increase between 500 and 700 fs, as indicated by the black arrows 1–3 in Fig. 2a. In addition, we observed characteristic modulations in the harmonic yield in the ranges of 100–400, 700–1000, and 1200–1600 fs. To extract the modulation frequencies, we applied a short-time Fourier transform with a 250-fs FWHM Gaussian window to the raw data (red dots in Fig. 2a). As shown in Fig. 2c, the spectrogram shows that the modulation frequencies distinctively changed at 400, 700, and 1100 fs, i.e., at the delay times when the harmonic yield significantly changed, as described above. The observed other-order harmonics showed the same trend. For comparisons with different high-harmonic orders, please refer to the supplemental material [*Supplemental*].

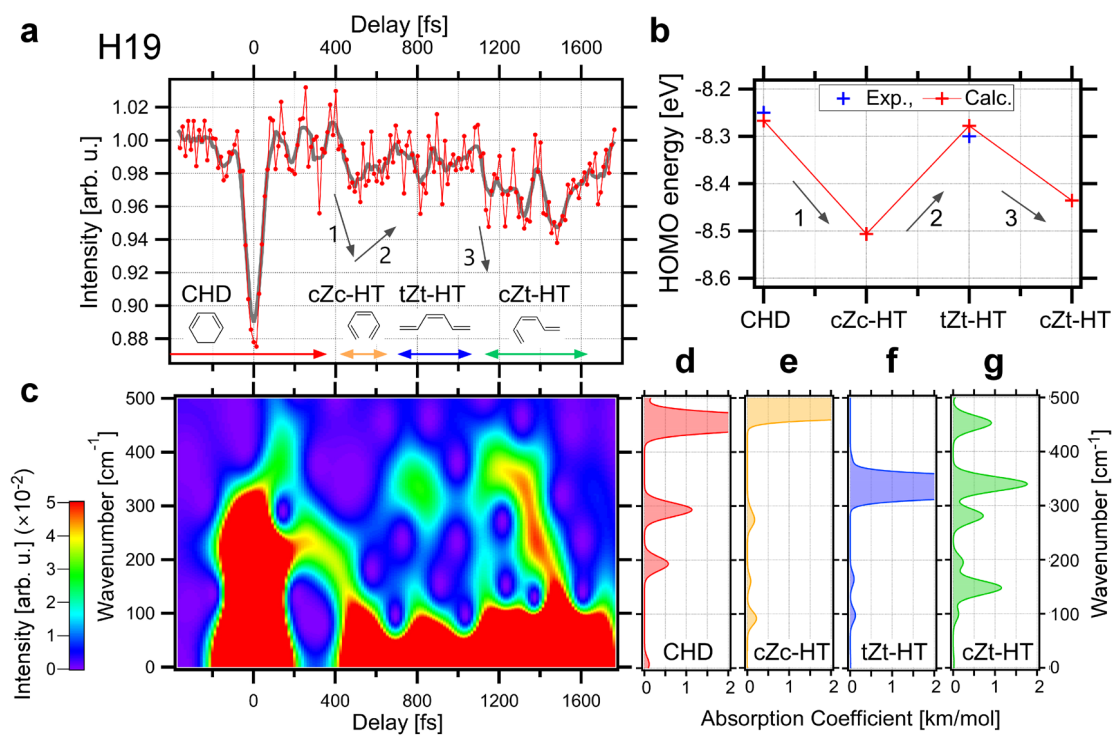

**Fig. 2** (a) Experimentally observed 19th harmonic (H19) yield as a function of pump-probe delay. Colored horizontal arrows denote evolution of the molecular structure. Red points denote raw data. The gray curve shows the smoothed data obtained by boxcar-averaging with a window size of five data points. (b) Calculated energies of the HOMOs of CHD and its isomers, cZc-, tZt-, and cZt-HT (red crosses). Blue crosses denote the experimentally observed first ionization energies of CHD[45] and tZt-HT[46], but with the opposite signs. (c) Short-time Fourier spectra of the transient harmonic yield [red curve in (a)]. (d–g) Calculated vibrational spectra of CHD and its isomers.

## Discussion

To associate the temporal evolution of the harmonic yield with the isomerization dynamics, we performed quantum-chemical calculations of the ground states of CHD and its isomers at the LC-BLYP[47]/cc-pVDZ[48] level using the GAMESS-package[49,50]. Figure 2b depicts the calculated HOMO energy levels of the isomers. For CHD and tZt-HT, the experimental values of the first ionization energies have been reported[45,46] (blue crosses in Fig. 2b; with their signs inverted). The calculated values agree very well with the experimental values. We also calculated the vibrational modes of these isomers, and the results are shown in Figures 2d–g for CHD, cZc-, tZt-, and cZt-HT, respectively. We multiplied the calculated vibrational frequencies by a scaling factor of 0.97 to be consistent with the experimental results for CHD[51]. Furthermore, we convolved a 30-cm$^{-1}$ FWHM Gaussian with the calculated spectra. For comparison with vibrational modes having wavenumbers larger than 500 cm$^{-1}$, please refer to the supplemental material [*Supplemental*].

Here, we associate the sudden changes in the harmonic yield with changes in the HOMO energy levels due to isomerization. This is because the change in HOMO energy levels alters the ionization rate and hence the harmonic yield. Furthermore, we also associate the modulations of the harmonic yield with molecular vibrations that perturb the HOMO energy levels and hence the harmonic yield[17,18].

The isomerization processes discussed in the following paragraphs are graphically summarized in Fig. 2a (horizontal colored arrows). The electronic excited state of CHD almost completely relaxes to the ground state within 80 fs. When the excited state relaxes, the vibrational modes of CHD are coherently excited through internal conversion[52]. The CHD thus remains in a vibrationally excited but electronically ground state between 80 and 400 fs. The relaxation to the ground state of CHD is indicated by the return of the harmonic yield to unity, i.e., to the same value as before photoexcitation (Fig. 2a). If the CHD had stayed in the electronic excited state, the harmonic yield would have increased because the excited state has a smaller ionization energy and results in a higher ionization rate. Since the harmonic yield remains close to unity between 80 and 400 fs, we conclude that the contribution from the excited state of CHD was obscured by the coherent artifact around zero delay and that the CHD returns to its electronic ground state within 80 fs and remains there until 400 fs. Note that, although tZt-HT has a HOMO energy level close to that of CHD (Fig. 2b), the possibility of tZt-HT formation at this stage is excluded. As Fig. 2c shows, we observed vibrational modes at 190, 290, and 450 cm$^{-1}$ between 80 and 400 fs; these modes are peculiar to CHD (Fig. 2d) and not to tZt-HT (Fig. 2f). Therefore, we conclude that the photo-excited CHD returns to its ground state before the ring-opening.

At 400 fs, the harmonic yield decreased steeply (black arrow 1 in Fig. 2a). Since the larger ionization energy of cZc-HT results in a lower harmonic yield, the sudden decrease indicates ring-opening to form cZc-HT. That is, the $C_1$–$C_6$ bond of CHD (Fig. 1) breaks, and cZc-HT is formed. The vibrational modes also changed at 400 fs (Fig. 2c). The disappearance of the characteristic modes of CHD around 190, 290, and 450 cm$^{-1}$ supports the formation of cZc-HT, whose vibrational spectrum is shown in Fig. 2e. However, we did not observe the vibrational mode around 500 cm$^{-1}$ that appears in Fig. 2e between 400 and 700 fs (Fig. 2c). There are two possible reasons for this discrepancy: (i) the unobserved vibrational mode is not excited in the reaction path, and (ii) the unobserved vibrational mode hardly affects the HOMO energy level and hence the harmonic yield.

Between 700 and 1100 fs, the molecule takes the form of tZt-HT, as is supported by the appearance of a vibrational mode at 350 cm$^{-1}$ in Figs 2c and 2f. The gradual recovery of the harmonic yield to nearly unity between 400 and 700 fs, as indicated by black arrow 2 in Fig. 2a, also indicates the transformation from cZc-HT to tZt-HT. Since tZt-HT has a smaller ionization energy than cZc-HT, but not smaller than CHD (Fig. 2b), the harmonic yield increases along with rotations about the $C_2$–$C_3$ and $C_4$–$C_5$ bonds of cZc-HT, but it does not increase above unity.

At 1100 fs, the harmonic yield decreased steeply (black arrow 3 in Fig. 2a), and several vibrational modes appeared below 500 cm$^{-1}$ (Fig. 2c). The newly appearing vibrational modes are in good agreement with those of cZt-HT shown in Fig. 2g. The decrease in the harmonic yield due to the larger ionization energy of cZt-HT compared with that of tZt-HT also supports the isomerization from tZt-HT to cZt-HT at 1100 fs.

The delayed ring-opening of CHD until about 400 fs after 2PA is interesting to note. Ring-opening in less than 200 fs after 1PA has been reported in many previous studies[27,29–32,35,38,39]. In a previous investigation using TR-PES, we also observed delayed ring-opening after 2PA[28]. Here, we revisit the results of the TR-PES experiment to verify consistency with the results obtained above. Figure 3 compares the results of TR-HHS and TR-PES. Figure 3b shows the transient yields of photoelectrons from the molecular orbitals (MOs) related to the C–C and $CH_2$ bonds (red) and from the MOs related to the C=C bonds (blue). Figure 3c shows the difference between these photoelectron yields.

By comparison with the TR-HHS results, we can interpret the transient photoelectron yields in a more detailed way, as follows: Since the increase in the photoelectron-yield difference at around 400 fs (Fig. 3c) indicates a decrease in the number of C–C bonds and an increase in the number of C=C bonds, we ascribe this behavior to ring-opening[28]. Furthermore, the increase in this difference between 400 and 700 fs can now be attributed to isomerization from cZc-HT to tZt-HT, although we had previously considered it to be a sign of ring-opening from CHD to cZc-HT. The increase stops at around 700 fs, and the photoelectron yields (Fig. 3b) and their difference (Fig. 3c) subsequently remains nearly constant until ~1100 fs. This indicates that the molecule retains the same structure, i.e., tZt-HT, between 700 and 1100 fs. At around 1100 fs, the photoelectron yields start to decrease (Fig. 3b), indicating that further isomerization occurs around this moment. In contrast, the difference remains almost unchanged after 1100 fs (Fig. 3c), indicating that the molecule remains in one of the hexatriene structures. The observed timescales and interpretations of TR-PES[28] and those of TR-HHS are thus entirely consistent with each other.

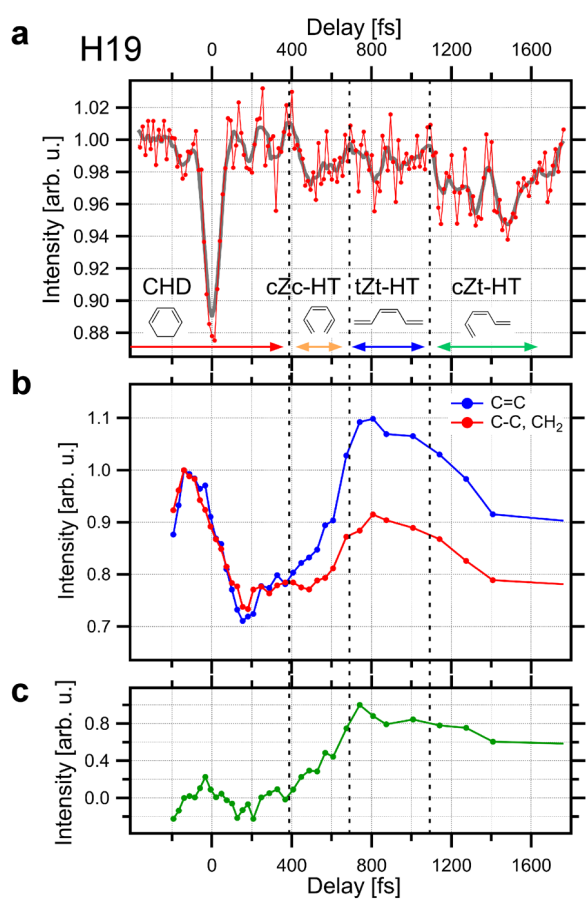

**Fig. 3** (a) Result of TR-HHS (identical to Fig. 2a). (b) Result of TR-PES (replot of Fig. 4c in ref. [28]). Red points depict the time-dependent yield of photoelectrons mainly originating from the MOs related to the C–C and $CH_2$ bonds of CHD. Similarly, blue points show the yield of photoelectrons from the C=C bonds. (c) Result of subtracting the red points from the blue points in (b).

In summary, we have shown that TR-HHS is a powerful tool for studying ultrafast photo-chemical reactions because of its sensitivity to both the electronic and nuclear dynamics. By associating the experimentally observed, transient high-harmonic yields with the calculated HOMO energies and vibrational spectra of the isomers, we have elucidated the photo-isomerization dynamics of CHD after 2PA of 3.1 eV photons. Furthermore, we have confirmed the consistency of these results with previously performed TR-PES measurements[28]. The question that remains is regarding the difference in the ring-opening dynamics of CHD upon 1PA and 2PA. We suspect that the differences in the initial photo-excited states and their relaxation pathways may be responsible for the different timescales of ring-opening. Future explorations of TR-HHS with higher time resolution may yield further insights into the initial dynamics of chemical reactions, i.e., the dynamics of electronic excitations in molecules, and so may answer this question.

## Methods

**Experimental details.**

Figure 4 is a schematic diagram of our experiment. A beam from a 1-kHz-repetition-rate Ti:sapphire chirped-pulse amplifier passes through a 500-μm-thick $LiB_3O_5$ crystal for second-harmonic generation. Then, it is split into two color channels using a dichroic mirror. The delay, power, polarization, and dispersion were separately adjusted. The second harmonic (3.1 eV, 10 μJ) was used to excite an electron in CHD from the HOMO to the $3p_x$-Rydberg state[34,53–57] *via* 2PA (pump), and we used the fundamental (1.55 eV, 730 μJ) to generate high harmonics (probe) from CHD. We recorded the high-harmonic yields as a function of the pump-probe delay. The relative polarizations of the pump and probe pulses were set to the magic angle (54.7°) to eliminate any time-dependent rotational effects. The separated beams were superimposed again using a dichroic mirror and collinearly focused using a concave mirror with a 500-mm focal length. The intensities of the pump and probe pulses in the interaction region were estimated to be 1.3 TW/cm$^2$ for a full-width at half-maximum (FWHM) pulse duration of 74 fs and 110 TW/cm$^2$ for a 30-fs FWHM pulse duration.

Liquid CHD (Sigma Aldrich, 97%), stored in a stainless-steel reservoir at room temperature, was continuously flowed into the vacuum chamber using 1.6-atm He gas through a 1-cm long, hollow glass fiber with an inner diameter of 100 μm. We avoided heating the sample to prevent thermal denaturation. The harmonics generated from CHD were spectrally separated using an EUV spectrometer and were imaged onto an EUV-sensitive charge-coupled device camera. Two 200-nm-thick aluminum filters were placed in front of the spectrometer to filter out the pump and probe light. We confirmed that harmonics from He were not observed by introducing He alone into the system.

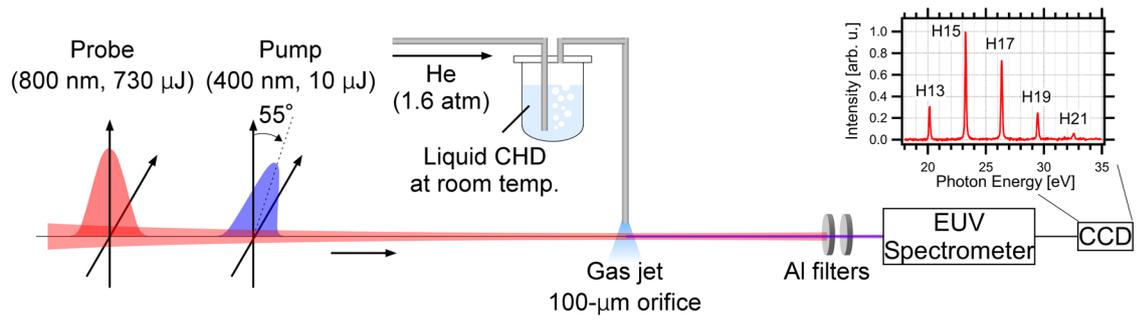

**Fig. 4** Schematic of the experimental setup. The inset shows the observed harmonic spectrum.

**Acknowledgments**

This work is supported by CREST, JST (JPMJCR15N1). T.S. is also supported by KAKENHI (15H03702 and 16K13854).

Supplementary materials for

# Time-resolved high-harmonic spectroscopy of ultrafast photo-isomerization dynamics

Keisuke Kaneshima, Yuki Ninota, and Taro Sekikawa

**Comparison between observations of different high-order harmonics**

Figure S1 compares the transient harmonic yields observed for different high-order harmonics. All observed harmonic orders are modulated in phase. This suggests that the ionization step contributes to modulating the high-harmonic signal because a modulation in the ionization rate would affect all orders, as discussed in[1]. In contrast, the variation of modulation was larger for higher-order harmonics. This can be attributed to the recombination of the electron with the parent molecule because higher harmonics are generated by recolliding electrons with shorter de Broglie wavelengths and they are more sensitive to small displacements of atoms[1]. Table S1 summarizes the de Broglie wavelengths of the observed harmonics, assuming that the ionization energy of 1,3-cyclohexadiene (CHD) is 8.25 eV[2]. The de Broglie wavelengths in our experiment are of the same order as the dimensions of CHD, as shown in Fig. S2. The higher-order harmonics that can be generated by longer-wavelength driver pulses are thus able to probe structural deformations with higher sensitivity[3].

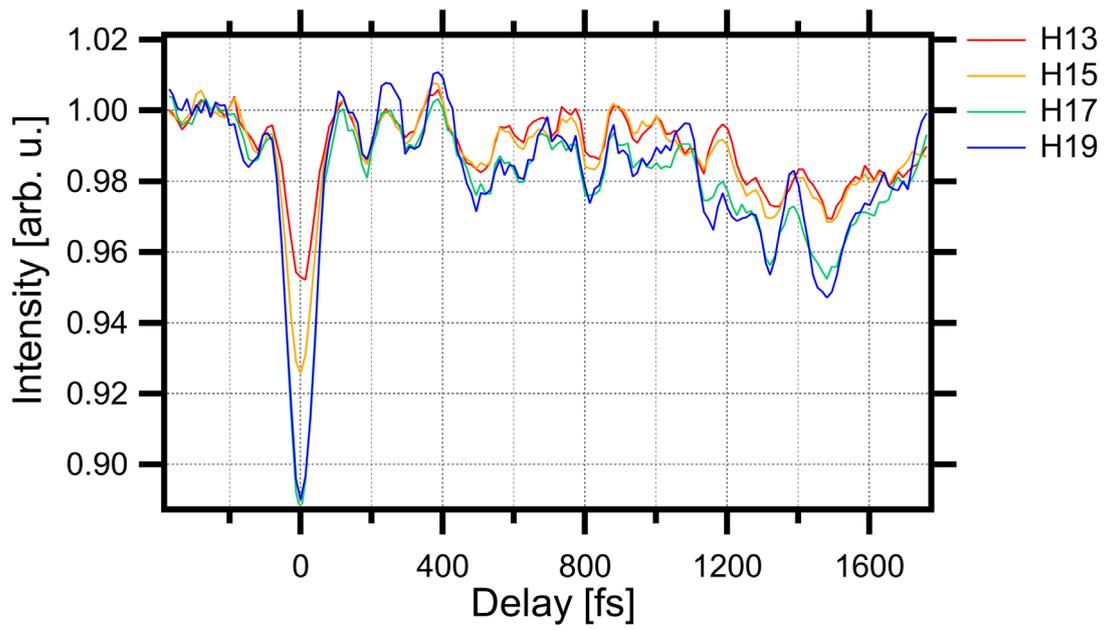

FIG. S1. Comparison of the transient harmonic yields observed for different high-order harmonics.

TABLE S1. de Broglie wavelengths of the observed harmonics.

|  | **H13** | **H15** | **H17** | **H19** |
|---|---|---|---|---|
| **de Broglie wavelength [Å]** | 3.55 | 3.16 | 2.87 | 2.66 |

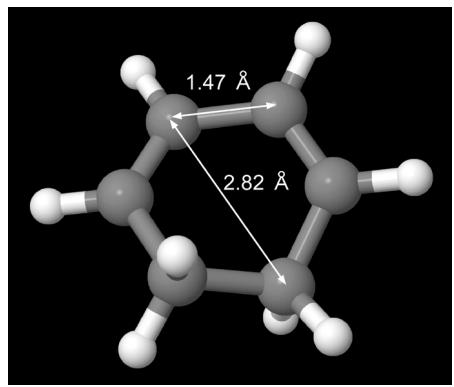

FIG. S2. Inter-atomic distances in CHD.

**Comparison of the experimental and calculated vibrational spectra**

Figure S3 is nearly identical to Fig. 2 but with an extended frequency range. In Fig. S3(b), the disappearance at ~400 fs delay of the ~650 cm$^{-1}$ component, which is a signature of CHD (Fig. S3c), strongly indicates that ring-opening occurs at this moment. The appearance of the 900–1000 cm$^{-1}$ components after ~400 fs also supports this explanation because these components are signatures of hexatriene conformers, as shown in Figures S3(c)–(f).

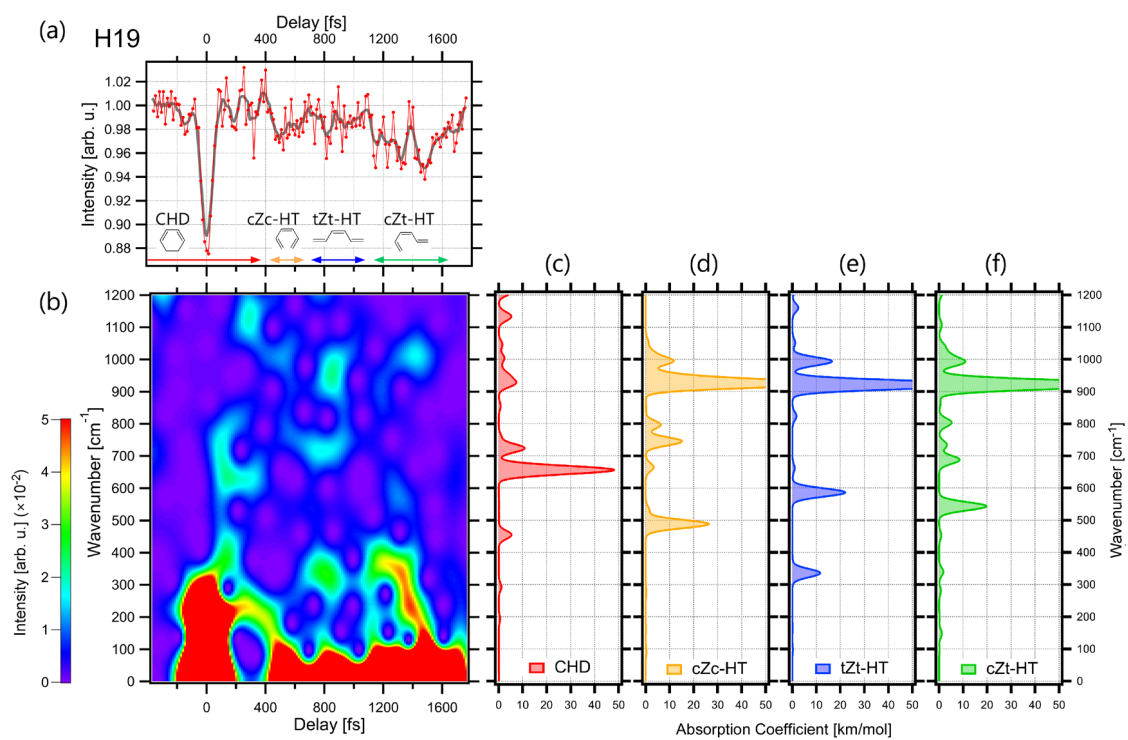

FIG. S3. (a) Experimentally observed transient 19th harmonic (H19) yield (identical to Fig. 2a). (b) Short-time Fourier spectra of the transient harmonic yield with a 250-fs full-width half-maximum (FWHM) Gaussian window (identical to Fig. 2c, but with an extended frequency range). (c–f) Calculated vibrational spectra of the isomers (identical to Figures 2c–f but with an extended frequency range).